\documentclass[12pt]{iopart}
\usepackage{graphicx}
\usepackage{dcolumn}
\usepackage{bm}

\begin{document}

\title[Local Inhomogeneity in Asymmetric Simple Exclusion Processes]{Local
Inhomogeneity in Asymmetric Simple Exclusion Processes with Extended Objects}

\author{Leah B. Shaw\dag\ddag, Anatoly B. Kolomeisky\P and Kelvin H. Lee\ddag}

\address{\dag\ Department of Physics, Cornell University, Ithaca, NY 14853-2501}

\address{\ddag\ School of Chemical and Biochemical Engineering, Cornell
University, Ithaca, NY 14853-5201}

\address{\P\ Department of Chemistry, Rice University, Houston, TX 77005}

\begin{abstract}
Totally asymmetric simple exclusion processes (TASEP) with particles which occupy more than one lattice site and with a local inhomogeneity far
away from the boundaries are investigated. These non-equilibrium processes are relevant for the understanding of many biological and chemical
phenomena. The steady-state phase diagrams, currents, and bulk densities are calculated using a simple approximate theory and
extensive Monte Carlo computer simulations. It is found that the phase diagram for TASEP with a local inhomogeneity is qualitatively similar to
homogeneous models, although the phase boundaries are significantly shifted. The complex dynamics is discussed in terms of domain-wall theory for
driven lattice systems.

\end{abstract}


\ead{tolya@rice.edu}

\submitto{\JPA}

\maketitle

\section{Introduction}

In recent years, asymmetric simple exclusion processes (ASEP) have
become a subject of increasing  scientific interest because of
their crucial role in the  investigations of numerous  dynamic
phenomena in chemistry, physics and biology
\cite{derrida98,schutz}. They are important for understanding
mechanisms of biopolymerization \cite{macdonald}, reptation
polymer dynamics \cite{schutz99}, diffusion through membrane
channels \cite{chou98}, traffic problems \cite{popkov01}, and protein
synthesis \cite{macdonald,lakatos03,shaw03}. ASEP are
non-equilibrium one-dimensional lattice models where particles
interact only through hard-core exclusion and move preferentially
in one direction.

Although dynamic rules of asymmetric simple  exclusion processes
are very simple, they show a very rich dynamic phase behavior. One
of the most striking features of ASEP with open boundaries is the
occurrence of non-equilibrium  phase transitions between
stationary states that have no analogs in equilibrium systems
\cite{derrida98,schutz,kolomeisky98}. The physics of these phase
transitions can be explained by utilizing a   phenomenological
domain-wall theory \cite{kolomeisky98}. According to this theory,
the entrance, the bulk, and the exit of the system define
their own stationary domains with  specific uniform densities and
currents. Domain walls exist in the border region between
different domains, and  the dynamics of these  domain walls,
which depends on the system parameters, in the stationary-state
limit  will determine the dominant phase. In the maximal-current
phase, the steady-state density profile and current are due to the
bulk dynamics, while the low-density (high-density) phase is enforced by
the entrance (exit) rate.

Recently, a new class of ASEP with particles occupying more than one lattice site  has been investigated using various mean-field and continuum
approaches \cite{lakatos03,shaw03}. These models provide a more realistic description of many biological processes such as RNA translation,
vesicle locomotion along filaments, and proteins sliding along DNA. For example, in RNA translation a ribosome typically covers 10-12 codon sites
but moves only one codon at a time \cite{stryer}. However, one more important dynamic feature, the non-uniformity of hopping rates, should also be
considered for ASEP with extended objects in order to give a  more realistic description of one-dimensional biological transport phenomena.
This issue has been only briefly considered by using crude approximation  methods and Monte Carlo simulations \cite{shaw03}. More thorough
theoretical investigations of effects of inhomogeneity  are needed. The main goal of this paper is to develop  a theoretical  and
computational description  of inhomogeneous ASEP  with particles of arbitrary size. We consider here the totally asymmetric simple exclusion
processes (TASEP), where particles can only move in one direction, although our theoretical arguments and our method can be extended to more
general exclusion models.

We present a theoretical study of the simplest inhomogeneous asymmetric exclusion model with particles of arbitrary size. In this  model,
the only nonuniform hopping rate is at the site which is in the middle of the lattice. We investigate this system by using a simple approximate
theory, based on the domain-wall approach, and extensive computer simulations.  The paper is organized as follows. In section 2 we outline our
model and known results for homogeneous ASEP with extended objects, and we present calculations from a simple approximate theory and domain- wall
arguments. Monte Carlo simulations are discussed in section 3. We summarize and conclude in section 4.

\section{TASEP with extended objects}

\subsection{Model}

\begin{figure}[tbp]
\centering
\includegraphics[clip=true]{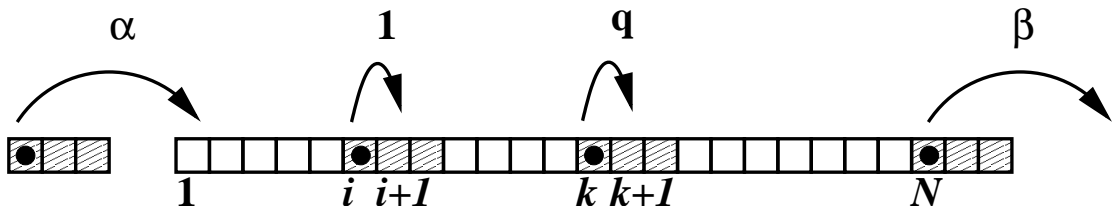}
\caption{Model for TASEP with extended objects and a local inhomogeneity.  System with size $l=3$ particles is shown.  Hopping rate in the bulk is
$1$.  Hopping rate at special site $k$ is $q$.  Entrance and exit rates are $\alpha$ and $\beta$, respectively.}
\label{fig:modelpic}\end{figure}

We consider identical  particles moving on a one-dimensional lattice consisting of $N$ sites, as shown in Fig. \ref{fig:modelpic}. Each particle
may cover $l \geq 1$ sites. For convenience, we associate the position of each particle with the position of its left edge. In the bulk of the
system, a particle located at site $i$ will move to the next site $i+1$ with the rate 1, given that site $i+l$ is empty. Furthermore, when
the particle is at the special site $k$, far away from the boundaries, the particle jumps forward to site $k+1$ with the rate $q$
(provided that site $k+l$ is empty). Thus particles travel from the left to the right with uniform rates except at the special site $k$, which is
the place of local inhomogeneity. Particles also can enter the system from the left with rate $\alpha$ if the first $l$ sites are empty.
When a particle reaches site $N$, it can exit with  rate $\beta$. We consider $N$ to be very large, and in this thermodynamic limit the
exact details of entrance and exit dynamics are not very important \cite{lakatos03}.  As in \cite{shaw03}, we define the density of the particles
in the system as the coverage density, namely, for $M$ particles on the lattice the density is given by $\rho=M l/N$. Since we discuss only the
totally asymmetric simple exclusion model with extended objects,  particles can  move in only one direction, from left to right as shown in Fig.
\ref{fig:modelpic}.

\subsection{Results for homogeneous TASEP with extended objects}

Analytical mean-field calculations and domain-wall arguments, supported by extensive Monte Carlo simulations, indicate that homogeneous TASEP with
extended objects has the same  three-phase diagram as homogeneous standard TASEP with $l=1$ \cite{lakatos03,shaw03}.

For $\alpha \geq \frac{1}{\sqrt{l}+1}$ and $\beta \geq \frac{1}{\sqrt{l}+1}$, the dynamics of the system is determined in the bulk,  and we
have a maximal-current phase with the stationary current and bulk density given by

\begin{equation}
J = \frac{1}{(\sqrt{l}+1)^{2}}, \quad \rho=\frac{\sqrt{l}}{\sqrt{l}+1}. \label{maxcurrent}
\end{equation}

When the particle entry is the rate-limiting step, which is realized for $\alpha < \frac{1}{\sqrt{l}+1}$ and $\alpha <\beta$, the system is in a
low-density stationary phase with the following current and bulk densities:

\begin{equation}\label{ld}
J=\frac{\alpha(1-\alpha)}{1+(l-1)\alpha}, \quad \rho=\frac{\alpha l}{1+ (l-1) \alpha}.
\end{equation}

The conditions $\beta < \frac{1}{\sqrt{l}+1}$ and $\beta <\alpha$ induce a high-density phase, which is governed by the exit dynamics. In
this case the current and bulk densities are given by

\begin{equation}\label{hd}
J=\frac{\beta(1-\beta)}{1+(l-1)\beta}, \quad \rho=1-\beta.
\end{equation}

There are two types of phase transitions in this system. A
first-order phase transition, with discontinuous change in density, is observed
when $\alpha=\beta$. However, the transitions between low-density
(high-density)  and maximal-current phases are continuous.

\subsection{Approximate solutions of inhomogeneous TASEP with extended objects}

For simplicity, let us assume that the size of the system $N$ is a
large even number and that the special site with jumping rate $q$ is
positioned at $k=N/2$. The exact position of local inhomogeneity
is not important as long as it is  far away from both boundaries
\cite{kolom98,mirin03}.

The special site $k$ breaks the translational symmetry of original homogeneous TASEP.  However, it also divides the system of size $N$ into two
homogeneous translationally invariant sublattices of size $N/2$. This observation allows us to consider our model with local inhomogeneity as two
coupled homogeneous TASEP with extended objects. The sublattices are coupled by a condition that the stationary currents in the left and right
subsystems should be the same. As a result, to calculate the properties of TASEP with extended objects and with local inhomogeneity, we can
use the known results for homogeneous TASEP. This approach has been used successfully before in different inhomogeneous asymmetric exclusion
models \cite{kolom98,mirin03}. Note that the results of Ref. \cite{kolom98} correspond to our special case $l=1$, i.e., when each particle
occupies only one site.

The main assumption of our theoretical method is that TASEP with local inhomogeneity can be viewed as two {\it independent} homogeneous TASEP
coupled only by the requirement to have the same steady-state  currents. Then each of the sublattices may exist in one of three different
stationary states, and  there are nine possible phases in the overall system. However, since the particle current in the left
subsystem should be equal to the current in the right subsystem, it is impossible to have the maximal-current phase in one of the
sublattices while the other sublattice is in the low-density or high-density phase. This observation eliminates four possible stationary
phases and leaves only low-density/low-density (ld/ld), high-density/high-density (hd/hd), low-density/high-density (ld/hd),
high-density/low-density (hd/ld) and maximal-current/maximal-current (mc/mc) stationary phases. In this notation, the first term corresponds to
the state of the left sublattice, while the second term describes the right sublattice.

Consider in a more detail the possibility of existence of an ld/hd phase in our system. This phase may exist only when $\alpha \neq \beta$
and
\begin{equation} \label{ld/hd}
\alpha < \frac{1}{\sqrt{l}+1}, \quad  \beta <\frac{1}{\sqrt{l}+1}.
\end{equation}
Then, because the currents in the sublattices are the same,
\begin{equation}
J=\frac{\alpha(1-\alpha)}{1+(l-1)\alpha}=\frac{\beta(1-\beta)}{1+(l-1)\beta},
\end{equation}
which yields
\begin{equation}
\beta =\frac{1-\alpha}{1+\alpha(l-1)}.
\end{equation}
However, using the condition (\ref{ld/hd}) for $\alpha$ we may conclude that $\beta
>\frac{1}{\sqrt{l}+1}$, which violates the condition for existence of the  high-density phase on the right sublattice. Thus the ld/hd phase also cannot be realized in our system. 

The local inhomogeneity in our model is realized for different
values of $q$. When $q>1$ the fast jumping rate is introduced.
In this case, the dynamics near the local inhomogeneity
will not affect the overall dynamics in the system since crossing
the special site $k$  will not be a rate-limiting step. As a
result, we have exactly the same phase diagram as for homogeneous TASEP
with extended objects  with ld/ld, hd/hd and mc/mc phases. The
presence of local inhomogeneity will only modify the density
profiles near the site $k$.

The more interesting case is $q<1$, when the dynamics near local inhomogeneity may determine the overall behavior of the system. Now consider in
more detail the process of a particle crossing from the left sublattice to the right one. The particle first approaches the right sublattice
when its left edge is on the site $k-l+1$. Then it makes $l-1$ jumps with rate 1 and one jump with the rate $q$, if these motions are allowed, and
the particle transfers  completely into the right subsystem. Thus, the effective rate of moving from the left sublattice to the right sublattice
is given by
\begin{equation}\label{qeff} q_{eff} =\frac{l}{1/q+(l-1)}=\frac{lq}{1+q(l-1)}.
\end{equation}

If the entrance to the left subsystem is the rate-limiting step,
then the system will be in the ld/ld phase with the particle current
and bulk density  given in (\ref{ld}). This situation is
realized for $\alpha$  less than some $\alpha^{*}$. Similarly,
when the exit from the right subsystem determines the overall
dynamics, the hd/hd phase exists  with the particle current
and bulk density given in (\ref{hd}), again for $\beta<\beta^{*}$.
Parameters $\alpha^{*}$ and $\beta^{*}$ will be calculated
explicitly below.

When the transition from the left sublattice to the right sublattice becomes the rate-limiting step, the hd/ld phase is realized in our system.
The parameters for existence of this phase can be found from the condition that the current through left sublattice is equal to the current
through the right sublattice and equal to the current passing through the local inhomogeneity. Following the  expressions (\ref{ld}) and
(\ref{hd}), the stationary current and bulk density in the left sublattice are given by
\begin{equation}\label{left}
J_{left}=\frac{\beta_{eff}(1-\beta_{eff})}{1+\beta_{eff}(l-1)}, \quad  \rho_{left}=1-\beta_{eff}, \end{equation} while in the right
sublattice,
\begin{equation}\label{right}
J_{right}=\frac{\alpha_{eff}(1-\alpha_{eff})}{1+\alpha_{eff}(l-1)},
 \quad  \rho_{right}=\frac{ l \alpha_{eff}}{1+\alpha_{eff}(l-1)},
\end{equation}
where $\beta_{eff}$ is the effective rate to exit from the left
sublattice and $\alpha _{eff}$ is the effective rate to enter the
right sublattice. The current passing through the local inhomogeneity
can be written as
\begin{equation}\label{j0} J_{0}=q_{eff}
\frac{\rho_{left}}{l}\frac{(1-\rho_{right})}{(1-\rho_{right}+\rho_{right}/l)}.
\end{equation}
This expression can be understood as follows. The parameter $q_{eff}$, which is given explicitly in (\ref{qeff}), is an effective rate of crossing
the local inhomogeneity. The factor  $\frac{\rho_{left}}{l}$ is the probability to find a particle at site  $k-l+1$, i.e.,  the particle
is crossing  from the left sublattice to the right sublattice.  Finally, $\frac{(1-\rho_{right})}{(1-\rho_{right}
+\rho_{right}/l)}$ is the conditional probability for the corresponding sites
in the right sublattice to be empty (as in \cite{shaw03}). The condition of stationary state  implies that $J_{left}=J_{0}=J_{right}$, which yields
\begin{eqnarray}
\rho_{right}&=&\frac{\rho_{left}-(1-q_{eff})[l-\rho_{left}(l-1)]}{1- (1-q_{eff})[l-\rho_{left}(l-1)]}, \nonumber \\
\rho_{right}&=&q_{eff}\rho_{left}.
\end{eqnarray}
Solving these equations leads to expressions for the densities in the
right and left sublattices:
\begin{eqnarray}\label{hd/ld-results}
\rho_{left}&=&\frac{(1+q_{eff})l}{2q_{eff}(l-1)}
\left[1-\sqrt{1-\frac{4q_{eff}(l-1)}{l(1+q_{eff})^{2}}}\right], \nonumber \\
\rho_{right}&=&q_{eff}\rho_{left}.
\end{eqnarray}

The explicit formula for the current can be found from (\ref{j0}). This
phase can only exist when $\alpha > \beta_{eff}$ and $\beta > \alpha_{eff}$,
and using (\ref{left}), (\ref{right}) and (\ref{hd/ld-results}) we
obtain that hd/ld phase is stable for $\alpha > \alpha^{*}$ and
$\beta > \beta^{*}$, where

\begin{equation}
\alpha^{*}=\beta^{*}=1-\frac{(1+q_{eff})l}{2q_{eff}(l-1)}\left[1-\sqrt{ 1-\frac{4q_{eff}(l-1)}{l(1+q_{eff})^{2}}}\right].
\end{equation}

Our theoretical results can be easily checked in some special limiting cases. For $q=q_{eff}=1$, and $\alpha>\alpha^*, \beta>\beta^*$, we obtain
that the bulk densities at the left and right sublattices are the same and equal to $\rho_{left}=\rho_{right}=\sqrt{l}/(\sqrt{l}+1)$, i.e., the
system is in the maximal-current phase, in agreement with known results \cite{lakatos03,shaw03}. In the limit of $q \rightarrow 0$, the effective
rate of crossing the local inhomogeneity is also very small, $q_{eff} \approx lq$. In this limit the bulk densities reduce to $\rho_{left}
\approx 1-q$ and $\rho_{right} \approx l q$, in agreement with intuitive expectations.  These results can be understood as follows. Particles will move into the right sublattice with a rate of $q$, so that the spacing between the particles will be about $1/q$ (distance
traveled before the next particle comes into the sublattice), giving a coverage density of
$lq$.  Now consider the left sublattice.  It will be nearly full, so consider the backwards current of holes moving through the slow site.  Holes
will enter the left sublattice at a rate $q$, so the hole density will be $q$.  Hence the coverage density in the left sublattice will be $1-q$.
 Another interesting limit is $ l
\gg 1$, which corresponds to the case of very large particles. In this limit we obtain $J \approx 1/l$, as was found before for the
homogeneous model \cite{lakatos03}.

It is interesting to note that our theory predicts a ``mixed'' nature of phase transitions from  hd/ld to ld/ld or hd/hd phases. At one sublattice
the transition will be continuous, while at the other  sublattice there is a jump in  density, which corresponds to  a first-order phase
transition.

\section{Monte Carlo simulations and discussions}

Our theoretical approach gives correct results in limiting cases. However,  to check the overall validity of our approximate theory we
performed Monte Carlo computer simulations.

Monte Carlo simulations were implemented similarly to those in \cite{shaw03}. Random sequential updating was used. Suppose that at some time the
lattice contains $M$ particles. Then in the next Monte Carlo step (MCS), $M+1$ particles are chosen at random, in sequence, to attempt moves. They
are selected from a pool containing the $M$ particles on the lattice plus a free particle that can enter the system with probability $\alpha$ if
the first $l$ sites are empty. Particles on lattice site $k$ advance with probability $q$ if site $k+l$ is empty.  Particles on site $N$ leave the
system with probability $\beta$.  All other particles advance if they have room to move.  Simulations were begun with an empty lattice and run
until steady state was reached. For particles with $l=12$, the size studied in greatest detail, the system size was $N=2500$, and $3\times
10^5$ MCS were allowed to reach steady state.  For other particle sizes, the system was run for at least $100 N$ MCS to reach steady state. After
steady state was attained, systems were simulated for an additional $1.5 \times 10^6$ Monte Carlo steps, during which the current leaving the
system was recorded constantly, and the particle density at each lattice site was sampled every 100 MCS.  Continuous time Monte Carlo
\cite{bortz75} was used to reduce the computation time required.

\begin{figure}[tbp]
\includegraphics[clip=true]{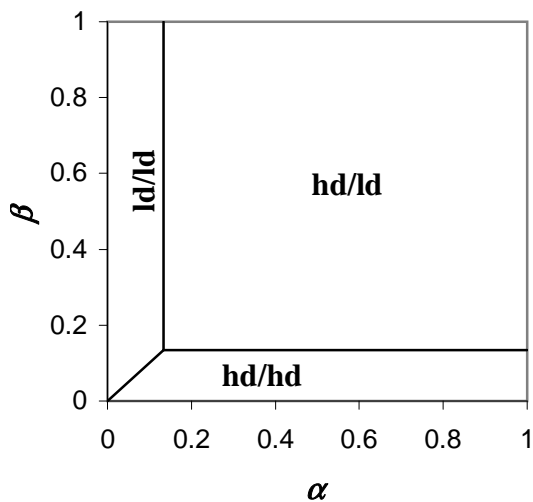}
\caption{Phase diagram for $l=12$, $q=0.2$.  Locations of phase transitions were determined via simulations to within $0.02$ and were in agreement
with predictions.}
\label{fig:phasediagram}
\end{figure}

\begin{figure}[tbp]
\includegraphics[clip=true]{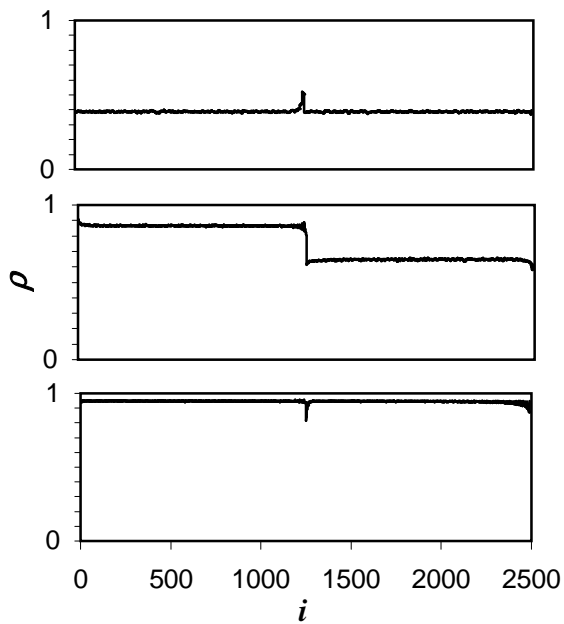}
\caption{Sample density profiles from simulations for the ld/ld phase (top, $\alpha=0.05, \beta=0.5$), the hd/ld phase (middle, $\alpha=0.5,
\beta=0.5$), and the hd/hd phase (bottom, $\alpha=0.5, \beta=0.05$).  All cases have $q=0.2$, $l=12$, $N=2500$.} \label{fig:profiles}
\end{figure}

\begin{figure}[tbp]
[a]\includegraphics[clip=true]{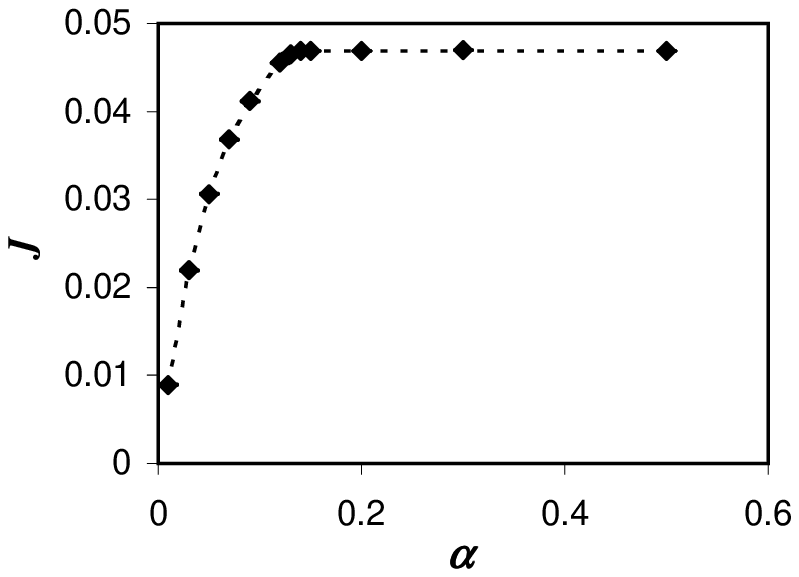}
[b]\includegraphics[clip=true]{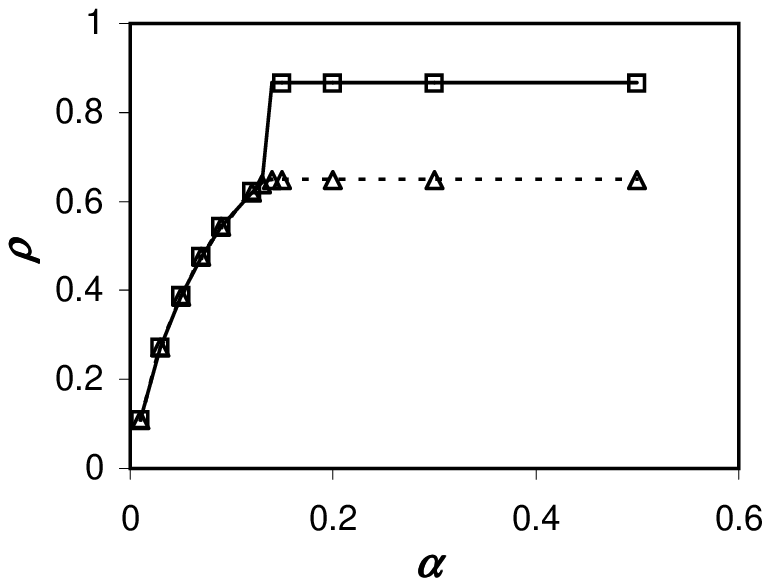}
\caption{(a) Dependence of current
$J$ on $\alpha$. (b) Dependence of bulk densities $\rho_{left}$ (squares, solid curve) and $\rho_{right}$ (triangles, dashed curve) on $\alpha$.
Bulk densities were spatial and time averages over regions in which the density profile was approximately constant. $\rho_{left}$ values are
omitted near the phase transition because the density profile was approximately linear in the left half of the system. In all cases, symbols are
simulation results, and curves are predicted values. Results are for $\beta=0.5, q=0.2, l=12, N=2500$. Error bars, determined from standard
deviations for simulations run in triplicate, are smaller than size of symbols.} \label{fig:current}
\end{figure}

The phase diagram was determined in detail from simulations for $l=12$, $q=0.2$, and $N=2500$ (see Fig. \ref{fig:phasediagram}). From the
theory, we expected the ld/ld phase to occur for $\alpha <\alpha^*=0.1338$, $\alpha<\beta$, the hd/hd phase to occur for $\beta<\beta^*=0.1338$,
$\beta<\alpha$, and the hd/ld phase to occur for $\alpha,\beta>0.1338$. Simulations were consistent with these expectations.  Sample density
profiles from each phase are shown in Fig. \ref{fig:profiles}.  The theoretical predictions for currents and bulk densities in each half of
the system also matched the simulations to within $ 1 \% $ (see Fig \ref{fig:current}).

\begin{figure}[tbp]
\includegraphics[clip=true]{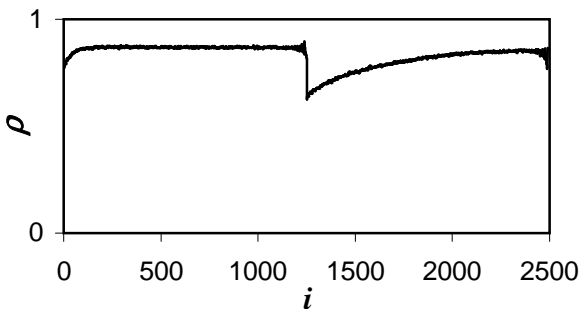}
\caption{Sample density profile near the transition from the hd/hd to the hd/ld phase.  Parameters are $\alpha=0.2,q=0.2,\beta=0.13,l=12,N=2500$.
The transition is expected to occur at $\beta=0.1338$.  Because of the proximity to the phase transition, the density profile is approximately
linear in the right half of the system.} \label{fig:transition}
\end{figure}

The phase transitions were identified as follows.  Near the transition between the hd/hd phase and the hd/ld phase, in the right half of the
system, $\alpha_{eff} \approx \beta$. This situation is expected to produce shock waves in the right half of the system (see \cite{shaw03}). We
detected these shock waves in simulations by observing the approximately linear density profile that they produce in the right half of the system.
An example of a density profile for the hd/hd to hd/ld transition is shown in Fig. \ref{fig:transition}.  Similarly, the left half of the
system exhibits an approximately linear density profile near the ld/ld  to hd/ld transition, and both halves of the system have approximately
linear density profiles near the ld/ld to hd/hd transition.  Visual inspection of the density profiles allows the transition lines to be localized
to within about $0.02$ units. Other techniques would be required to find the transition lines more precisely. For example, we could determine the
boundaries of the hd/ld phase by measuring the current as a function of $\alpha$ or $\beta$ and determining at which $\alpha$ or $\beta$ the
current saturates.

Simulations were conducted in each phase for other values of $l$ ($l=2, 5, 20$) to confirm the theoretical predictions. System sizes $N$ were
always at least $200 l$, in keeping with the assumption of large $N$.  The theoretical predictions for the current and bulk densities were
generally more accurate for larger particles. The greatest discrepancies between theory and simulations were seen for $l=2$ in the  hd/ld phase,
but observed discrepancies were always less than $10 \% $. These results show that our approximate theory  becomes more exact for larger-size
particles. This is because the particle-particle correlations decrease with increasing $l$. Consider a lattice consisting of $N=nl$ sites with
$M$ particles (each of size $l$). If we assume that particles are uniformly distributed, we can easily estimate the average number of empty sites
$x_l$ between two neighboring particles:  $x_{l}=\frac{nl-Ml}{M}=l x_{1}$. Thus the number of empty sites between the particles grows linearly with the particle size $l$. As
result, particles correlate to lesser degree with each other, and our theoretical mean-field approach becomes more accurate. Similar conclusions
have been reached in Ref. \cite{lakatos03}.

The phase diagram and phase transitions in TASEP with extended objects and with a local inhomogeneity can also be easily explained with the help
of the domain-wall theory \cite{kolomeisky98}. As we discussed above, the introduction of local inhomogeneity divides the system into the two
homogeneous sublattices, and the phase behavior in each sublattice is determined by the dynamics of domain walls. Thus there are always  two
domain walls  present in the system. In the hd/ld phase at stationary state the domain walls are localized near the entrance and the exit of the
system, while in ld/ld stationary  phase the domain walls can be found diffusing near the local inhomogeneity and the exit. Similarly, in hd/hd
stationary phase the domain walls are near the entrance and the local inhomogeneity. The phase transitions can be associated with the change of
position of domain walls. For example, the transition from ld/ld phase to hd/ld phase (when $q<1$) is the consequence of  the motion of the domain
wall from the position of local inhomogeneity to the entrance, while the second domain wall stays at the exit. This observation also
explains the ``mixed'' nature of this phase transition.

\section{Summary and conclusions}

We investigated the effect of  inhomogeneity in totally asymmetric simple exclusion processes with particles of arbitrary size. Specifically, the
simple model with one inhomogeneous jumping rate in the middle of the system was considered. The model was solved using a simple approximate
method. Our analytical approach was based on two assumptions. The first one is the fact that the local inhomogeneity divides the system into the
two coupled homogeneous systems, for which  very precise mean-field  solutions already exist. In the second approximation we assumed that the
dynamics of particles in each sublattice is independent, i.e., we  neglected the correlations between the particles in the two coupled subsystems.
Using these assumptions, we calculated explicitly the phase diagrams, bulk densities, and stationary currents.

Our theoretical results indicate that for a fast jumping rate ($q \geq 1$), the phase diagrams, currents and bulk densities do not change as
compared with homogeneous TASEP with extended objects. However, for the slow jumping rate, the situation is different. As in the
homogeneous TASEP, three stationary phases are found, although phase  boundaries, stationary currents and bulk densities change significantly.
Our theoretical predictions are well supported by extensive Monte Carlo simulations. Some small deviations between the analytical theory and
computer simulations are due to the neglect of correlations near local inhomogeneity  in our theoretical approach. However, the precision of
our theoretical predictions increases with the size of the particles. The microscopic nature of phase diagram was  analyzed using phenomenological
domain-wall theory.

Although we considered only totally asymmetric exclusion processes, our theoretical arguments can be easily extended to more general
partially asymmetric exclusion processes where particles can hop in both directions. However, the important parameter of effective crossing
rate from the left sublattice to the right should be calculated using more general expressions, $q_{eff}=l/\tau$, where $\tau$ is mean
first-passage time for the particle to move to the right sublattice. It can be calculated using standard procedures \cite{vanKampen}.

\ack
ABK acknowledges  support from the Camille and Henry Dreyfus New Faculty Awards Program (under Grant No. NF-00-056), from the Welch
Foundation (under Grant No. C-1559), and from the US National Science Foundation through the grant CHE-0237105. KHL acknowledges support from
the NSF through grant BES-0120315. LBS was supported by a Corning Foundation Fellowship and IBM Fellowship. This research was conducted using the
resources of the Cornell Theory Center, which receives funding from Cornell University, New York State, federal agencies, foundations, and
corporate partners.

\end{document}